\input epsf
\documentclass[nohyper,notoc]{article} 
\usepackage{color}
\usepackage{epsfig}
\usepackage{bm}

\usepackage[utf8]{inputenc} 
\usepackage[T1]{fontenc}

\def\ben{\begin{enumerate}}
\def\een{\end{enumerate}}
\def\bit{\begin{itemize}}
\def\eit{\end{itemize}}
\def\beq{\begin{equation}}
\def\eeq{\end{equation}}
\def\bea{\begin{eqnarray}}
\def\eea{\end{eqnarray}}
\def\bq{\begin{quote}}
\def\eq{\end{quote}}
\def \lsim{\mathrel{\vcenter
     {\hbox{$<$}\nointerlineskip\hbox{$\sim$}}}}
\def \gsim{\mathrel{\vcenter
     {\hbox{$>$}\nointerlineskip\hbox{$\sim$}}}}
\def\gappeq{\mathrel{\rlap {\raise.5ex\hbox{$>$}}
{\lower.5ex\hbox{$\sim$}}}}
\def\lappeq{\mathrel{\rlap{\raise.5ex\hbox{$<$}}
{\lower.5ex\hbox{$\sim$}}}}

\def\meg{\mu \to e \gamma}

\def\mec{\mu \! \to \! e~ {\rm conversion}}

\def\meee{\mu \to e \bar{e} e}

\def\a{\alpha}
\def\b{\beta}
\def\g{\gamma}

\def\m{\mu}

\definecolor{red}{rgb}{0,0,0}
\definecolor{blue}{rgb}{0,0,0}

\begin{document}

\renewcommand{\thefootnote}{\fnsymbol{footnote}}
\begin{center}
{\Large Selecting
$\mu \to e$  Conversion
 Targets to distinguish Lepton Flavour-Changing
 Operators
}

\vskip 25pt
{\bf    Sacha Davidson $^{1,}$\footnote{E-mail address:
s.davidson@lupm.in2p3.fr}   Yoshitaka Kuno  $^{2,}$
\footnote{E-mail address: kuno@phys.sci.osaka-u.ac.jp} 
and Masato Yamanaka $^{3,}$  }
\footnote{E-mail address: yamanaka@ip.kyusan-u.ac.jp} 
 
\vskip 10pt  
$^1${\it
LUPM, CNRS, Universit\'e Montpellier,
Place Eugene Bataillon, F-34095 Montpellier, Cedex 5, France
}\\
$^2${\it Department of Physics, Osaka University, 
1-1 Machikaneyama, Toyonaka, Osaka 560-0043, Japan}\\
$^3${\it Department of science and technology, 
Kyushu Sangyo University, Fukuoka 813-8503, Japan}

\vskip 20pt
{\bf Abstract}
\end{center}

\begin{quotation}
  {\noindent\small 
The experimental sensitivity to $\mu \to e$ conversion on nuclei is set to
improve by four orders of magnitude in coming years.
However, various operator coefficients 
add coherently  in the amplitude for $\mec$, weighted by
nucleus-dependent functions,
and therefore in the event of a detection, identifying
the relevant new physics scenarios could be difficult.
Using a representation of the nuclear targets as vectors in
coefficient space, whose components are the
weighting functions, we quantify the expectation
that different nuclear targets could give different
constraints.
{\color{blue}
We show that  all but two  combinations of the 10 Spin-Independent (SI)
coefficients could be constrained by future measurements,
but discriminating among the axial, tensor and pseudoscalar operators
that contribute to the Spin-Dependent (SD) process
 would require dedicated nuclear calculations.
 We anticipate that $\mec$ 
could constrain 10 to 14 combinations of coefficients;
if  $\meg$ and $\meee$ constrain eight more,  that  leaves
60 to 64 ``flat directions'' in the basis of  QED$\times$QCD-invariant
operators which describe $\mu \to e$ flavour
change below $m_W$.

}

\vskip 10pt
\noindent
}

\end{quotation}

\vskip 20pt

\section{Introduction}
\label{intro}

The observation of neutrino mixing and  masses implies that  flavour
cannot be conserved among charged leptons. However,
despite a  long programme of experimental searches
for various processes, charged lepton
flavour violation (CLFV)  at a point 
has yet to be observed. 

For $\mu \leftrightarrow e$ flavour change,
the  current  most stringent bound is $BR(\meg) \leq 4.2 \times 10^{-13}$
from the MEG collaboration \cite{MEG} at PSI.  
This sensitivity will
improve by one order of magnitude  in coming years \cite{MEG2}, 
and 
the Mu3e experiment \cite{Mu3e}  at PSI aims to reach
$BR(\meee) \sim 10^{-16}$. 
Several  experiments under
construction will improve the sensitivity
to $\mec$  on nuclei:
The COMET \cite{COMET} at J-PARC 
and the Mu2e \cite{Mu2e} at FNAL 
plan to reach branching ratios of $ BR( \mu Al \to e Al) \sim  10^{-16}$. 
The PRISM/PRIME proposal \cite{PP} aims to probe $\sim 10^{-18}$,
and at the same time enables to use heavy $\mec$ targets with shorter lifetimes 
of their muonic atoms, thanks to its designed pure muon beam with no pion contamination.
\footnote{Another interesting observable at these experiments is 
the $\mu^- e^- \to e^- e^-$ in a muonic atom. This process could 
have not only photonic dipole but also contact interactions, and the 
atomic number dependence of its reaction rate makes possible to 
discriminate the type of relevant CLFV interactions \cite{Koike:2010xr, 
Uesaka:2016vfy, Uesaka:2017yin}.} 
This enhanced sensitivity and  broader  selection of $\mec$ targets 
motivate  our interest in low-energy $\mu \leftrightarrow e$
flavour change. 

In the coming years,
irrespective of whether CLFV is observed
or  further constrained, it is
important to maximise the amount
of information that experiments can
obtain about the New Physics responsible for CLFV.
This is especially  challenging for the operators involving
nucleons or quarks, because
 in $\mu \to e$ conversion, the contributing coefficients add
in the amplitude.
So in this paper, we consider
$\mu \to e $ conversion on nuclei,  and present a
recipe for selecting targets such that they 
 constrain or measure different  CLFV parameters.
Reference \cite{CKOT}  is an earlier discussion of
the prospects  of distinguishing models with $\mec$.
{ A more recent publication~\cite{DKS} about
Spin-Dependent $\mu \to e $ conversion studied what could be learned
about  models or operator coefficients,  from
targets with and without spin. 
In this letter, we  follow the perspective of
\cite{DKS}, focussing on the Spin Independent  process,
and explore how many independent constraints
can be obtained on operator coefficients. }

We assume that the New Physics responsible 
for  $\mu \to e$
conversion is heavy,  and  parametrise it
 in Effective Field Theory (EFT) \cite{Georgi,LesHouches,KO,megmW}.
Section \ref{sec:2} gives  the  $\mec$ rate,
and the effective Lagrangian at the experimental scale  ($\sim $ GeV),
in terms of  operators   that are QED invariant,
labelled by their  Lorentz structure,
and  constructed  out of 
electrons, muons and nucleons ($p$ and $n$). 
In Section \ref{sec:3}  we divide   the rate into pieces
that do not interfere with each other.
Section \ref{ssec:picture} is  a toy model
of two observables that depend on a
sum of theoretical parameters, which illustrates the impact of
 theoretical uncertainties  on the determination of
 operator coefficients. It is well-known,
since the study of Kitano, Koike and Okada (KKO) \cite{KKO},
that different target nuclei have different relative
sensitivity to the various operator coefficients. 
In Section \ref{sec:5}, using the notion of targets
as vectors in  the space of operator coefficients
introduced in Reference \cite{DKS},
we  explore which current experimental bounds can
give independent constraints on operator coefficients,
given the current theoretical uncertainties.
Section \ref{sec:target} discusses the prospects  of future
experiments, {\color{blue} Section \ref{sec:FlatDir} compares the
number of  operator coefficients to the
number of constraints that could be obtained from
$\mec$, $\meee$ and $\meg$,} 
and Section \ref{sum} is the summary.


\section{ $\bm{\mec}$}
\label{sec:2}

$\mec$ is the process where an incident $\mu^-$ is captured by a
nucleus, and tumbles down to the $1s$ state. The muon can then
interact with the nucleus, by exchanging a photon or
via a contact interaction, and  turn into
an electron  which escapes with an energy $\sim m_\mu$.
This process has been searched for in the past
with various target materials, as summarised
in table \ref{tab:1}; the best existing bound
is $BR (\mu Au \to e Au) < 7 \times 10^{-13}$  on 
Gold ($Z=79$) 
from SINDRUM-II \cite{Bertl:2006up}.

The  interaction of the muon with the nucleus can
be parametrised at the experimental scale  ($\Lambda_{expt}$) in Effective Field Theory,
via dipole operators  and a variety of 2-nucleon
operators :
\bea 
{\cal L}_{\mu A \rightarrow  eA}(\Lambda_{expt})& =& 
-\frac{4G_F}{\sqrt{2}}   \sum_{N= p,n}  {\Big [}  m_\mu \left(
C_{D L} \overline{e_R} \sigma^{\a\b} \mu_L F_{\a\b} +
C_{D R} \overline{e_L} \sigma^{\a\b} \mu_R F_{\a\b}\right)
\nonumber\\
&&
+  \left( \widetilde{C}^{(NN)}_{SL}  \overline{e} P_L \mu
+  \widetilde{C}^{(NN)}_{SR} \overline{e} P_R \mu \right) \overline{N}  N
 \nonumber\\&&
  +   \left( \widetilde{C}^{(NN)}_{P,L}  \overline{e} P_L \mu
+  \widetilde{C}^{(NN)}_{P,R} \overline{e} P_R \mu \right) \overline{N}\g_5  N
 \nonumber\\&&
+   \left(   \widetilde{C}^{(NN)}_{VL} \overline{e}\gamma^{\a} P_L \mu
+  \widetilde{C}^{(NN)}_{VR} \overline{e}\gamma^{\a} P_R \mu \right) \overline{N} \g_\a N
\nonumber \\
&&
+  \left(   \widetilde{C}^{(NN)}_{A,L} \overline{e}\gamma^{\a} P_L \mu
+  \widetilde{C}^{(NN)}_{A,R} \overline{e}\gamma^{\a} P_R \mu \right) \overline{N} \g_\a\g_5 N
\nonumber \\
&&
+  \left(   \widetilde{C}^{(NN)}_{Der,L} 
\overline{e} \gamma^\a P_L \mu   +
\widetilde{C}^{(NN)}_{Der,R} 
\overline{e} \gamma^\a P_R \mu   \right)
i(\overline{N} \stackrel{\leftrightarrow}{\partial_\a} \g_5 N )
\nonumber \\
&&
+  \left(   \widetilde{C}^{(NN)}_{T,L} \overline{e}\sigma^{\a\b} P_L \mu
+  \widetilde{C}^{(NN)}_{T,R} \overline{e}\sigma^{\a \b} P_R \mu \right)
\overline{N} \sigma_{\a\b} N
 +h.c. {\Big ]}~~~.
 \label{CiriglianoN}
\eea
 Since the electron is relativistic, and
 the nucleons not, it is convenient to use
 a chiral basis for the lepton current, but
 not for the nucleons.

This  basis of nucleon operators
 is chosen  because it represents the minimal set
 onto which two-lepton-two-quark, and
 two-lepton-two-gluon operators can be matched at
 the leading order in $\chi$PT    
{\color{red} \footnote{At higher order in $\chi$PT,
additional operators can appear, sometimes involving
more than two nucleons \cite{HKS}.}. This explains the presence of
the derivative operators $\widetilde{{\cal O}}^{(NN)}_{Der,X}$,
which  
 represent pion exchange between the leptons
and nucleons at finite momentum transfer. 
They give a contribution to Spin-Dependent
$\mu \to e$ conversion that is comparable to the
 $\widetilde{{\cal O}}^{(NN)}_{A,X}$ operators \cite{DKS}.
We do not   count the coefficients
of the derivative operators as 
independent  parameters, because their effects
could be included as
a momentum-transfer-dependence of the
$G_A^{N,q}$ factors that relate quark
to nucleon axial operators \cite{DKS}.}

Like in WIMP scattering on nuclei, the muon
can interact coherently with the charge  or mass
distribution of the nucleus, called
the ``Spin Independent'' (SI) process, or it
can have  ``Spin-Dependent'' (SD)  interactions\cite{CDK}
with the nucleus at a rate that does not
benefit from the atomic-number-squared
enhancement of the SI rate. The Dipole,
Scalar and Vector operators will  contribute
to the SI rate (with a small admixture of the Tensor, see
eqn \ref{remplacer}), and the Axial, Tensor
and Pseudoscalar operators contribute
to the SD rate.

The SI contribution to 
 the branching ratio  for $\mec$  on the nucleus $A$
(${BR}_{SI}(A\mu \to Ae)$),
was calculated by 
Kitano, Koike and Okada (KKO) \cite{KKO} 
to be
\bea
{\rm BR}_{SI}(\mu A \to eA) &=&   \frac{32G_F^2 m_\m^5 }{ \Gamma_{cap}}     
 {\Big [ } \big|     
   \widetilde{C}^{pp}_{V,R} V^{(p)} + \widetilde{C}^{pp'}_{S,L}  S^{(p)}
+ \widetilde{C}^{nn}_{V,R} V^{(n)} + \widetilde{C}^{nn'}_{S,L} S^{(n)} 
+  C_{D,L} \frac{D}{4}  
 \big|^2   + \{ L \leftrightarrow R \}~ {\Big ]}~~,
\label{BRSIKKO}
\eea
where  $\Gamma_{capt}$ is
the rate for  the  muon to transform
to a neutrino by capture
on the nucleus~\cite{KKO,Suzuki:1987jf},
 {\color{red} ($= 0.7054 \times 10^6$/sec in Aluminium).}
The  nucleus ($A$) and nucleon($N\in\{n,p\}$)-dependent 
``overlap  integrals''   $D_A$, $S_A^{(p)}$,  $V_A^{(p)}$,
 $S_A^{(n)}$,  $V_A^{(n)}$,
correspond to the integral 
over the nucleus of the lepton wavefunctions
and the appropriate  nucleon density. These
overlap integrals will play a central role
in our analysis, and are 
given in KKO \cite{KKO}.
The primed scalar  coefficient includes
a small part of the tensor coefficient, because
the tensor contributes at finite momentum transfer to
the  SI process \cite{CDK,DKS}: 
\beq
\widetilde{C}_{{S},X}^{ NN'} =  \widetilde{C}_{{S},X}^{ NN}+ \frac{2 m_\mu}{m_N}
 \widetilde{C}_{T,X}^{NN} ~~~.~~~ 
\label{remplacer}
\eeq

The SD rate depends on the distribution of  spin in the nucleus, and
therefore requires  detailed nuclear modelling.
The tensor and axial vector contributions were estimated in  References 
\cite{CDK,DKS}
for light ($Z\lsim 20$) nuclei, where the muon wavefunction is wider than
the radius of the nucleus, and the electron can be
approximated as a plane wave.   In this limit, where
the muon wavefunction can be factored out of the
nuclear spin-expectation-value,  the nuclear calculation
of SD  WIMP scattering   on the  quark
axial current can be used  for $\mec$.  The SD branching ratio
 on a target $A$   of charge  $Z$,  with a fraction $\epsilon_I$ of
 isotope $I$ with spin $J_I$, can be estimated as
\bea
{\rm BR}_{SD}( \mu A \to  e A)\!\! &=& \!\! \frac{8G_F^2 m_\m^5(\alpha Z)^3 }{ \Gamma_{cap}\pi^2}    \left[
 \sum_{I}  4 \epsilon_{I}  \frac{J_{I}+1}{J_{I}}
\, \Big|   S^{I}_p (\widetilde{C}_{{A},L}^{pp}+ 2
 \widetilde{C}_{T,R}^{pp})
 + 
 S^{I}_n  (\widetilde{C}_{{A},L}^{nn}+ 2
 \widetilde{C}_{T,R}^{nn})
 \Big|^2    \  \frac{S_{I} (m_\mu)}{S_{I} (0)} 
+ \{ L \leftrightarrow R \} \right]
\nonumber\\
&&\label{BRCDK}
\eea
where $S_p^I$ is the proton spin  expectation value
in isotope $I$  at zero momentum transfer, and
$ {S_{I} (m_\mu)}/{S_{I} (0)} $ is a finite momentum
transfer correction, which  has been calculated
for the axial current in  some nuclei(see {\it eg}
References \cite{EngelRTO,KMGS}
for Aluminium; this factor includes
the derivative operators  $\widetilde{{\cal O}}^{(NN)}_{Der,X}$).
The targets which have been
used for $\mec$ searches are listed in  Table \ref{tab:1},
with the abundances  of some
spin-carrying isotopes, and  some results for  the
proton and neutron spin  expectation values. 

\begin{table}[th]
\begin{center}
\begin{tabular}{|l|c|c|c|c|c|c|}
\hline
target & isotopes[abundance] &J& $S_p^A$ , $S_n^A$ & $S_I(m_\mu)/S_I(0)$&$B_Z$& BR(90\% C.L.)\\
\hline
{Sulfur} &  Z=16,A= 32[95\%]&0& & &&  $ < 7 \times 10^{-11}$ \cite{Badertscher:1980bt}\\
\hline
{Titanium}& Z=22,A= 48[74\%]&0&&& 234&
 $ < 4.3\times 10^{-12}$\cite{Bertl:2006up}\\
& Z=22,A= 47[7.5\%] &5/2&0.0 , 0.21\cite{EVdeSim}&$\sim$0.12 && \\
 & Z=22,A= 49[5.4\%] &7/2&0.0 , 0.29\cite{EVdeSim}&$\sim$0.12 &&\\ 
\hline
{Copper }& Z=29,A= 63[70\%]& 3/2&& && $BR \leq 1.6 \times   10^{-8}$\cite{Bryman}
\\
& Z=29,A= 65[31\%]&3/2&& && \\
\hline
{Gold} &Z =79,A =197[100\%]&5/2& -(0.52$\to$0.30), 0.0 &&285&  $BR < 7\times 10^{-13}$\cite{Bertl:2006up}\\
\hline
{Lead}& Z=82, A=206[24\%] &0& && & $BR < 4.6\times 10^{-11}$  \cite{Bertl:2006up}\\
&Z=82, A=207[22\%] &1/2& 0.0 ,-0.15 \cite{EVdeSim}&0.55\cite{Bednyakov:2006ux}, $\sim$.026&&\\
& Z=82, A=208[52\%] &0& &&&\\ 
\hline
\hline
{Aluminium}  &  Z=13,A= 27[100\%]&5/2&
0.34, 0.030 \cite{EngelRTO,KMGS} 
&0.29 \cite{EngelRTO,KMGS}&132&  $ \to  10^{-16}$\\
\hline
\end{tabular}
\caption{Current  experimental bounds  on $\mec$ {\color{red}(the last line
gives the future
sensitivity on Aluminium), and parameters
relevant to  the SD calculation}. The isotope
abundances  are from \cite{AbondIso}. The parameter
$B_Z$ is defined in eqn (\ref{BZ}). The estimate for  $S_p^{Au}$ is 
based on the  Odd Group Model of \cite{EVdeSim}, assuming J=1/2.
 {\color{red} The estimated  form factors $S_I(m_\mu)/S_I(0)$
 for Titanium and Lead are  an
 extrapolation  from \cite{DKS}, discussed in the Appendix.}
\label{tab:1}}
\end{center}
\end{table}


\section{To determine or constrain  how many coefficients?}
\label{sec:3}

The Lagrangian of eqn (\ref{CiriglianoN})
contains twenty-two unknown operator coefficients
{\color{red} (not counting the derivative operators
as discussed after eqn (\ref{CiriglianoN}))}.
These coefficients contribute to various
observables, so can be constrained,
or measured, in  different ways:
\ben
\item   we neglect  
the two  dipole coefficients, because
 the upcoming MEG II and Mu3e experiments, 
 respectively searching for  $\meg$ and  $\meee$,
have a slightly better sensitivity:
if  MEG II  and Mu3e  set  bounds   $BR (\meg) < 6 \times 10^{-14}$ 
 and $BR(\meee) < 10^{-16}$, this would translate to 
$|C_{D,X}| \leq 2.0\times 10^{-9}$ (see eqn \ref{bdmegmeee}).
Whereas a  SI $\mec$  branching ratio of $10^{-16}$ on Aluminium
can  be sensitive  to $|C_{D,X}| \gsim 3.1\times 10^{-9}$.
\item the remaining  20  coefficients involving
nucleons can be divided into two classes,
labelled by the chirality/helicity of
the outgoing (relativistic) electron. 
The interference between these classes
 is usually neglected (suppressed
 by  $m^2_e/m^2_\mu$), so an experimental upper bound
 on the rate  simultaneously sets bounds
on the coefficients of both chiralities.
If a $\mec$ signal is observed with polarised
muonic atoms, it could
be possible to identify the chirality of
the operator by measuring an angular distribution
of the electron with respect to the muon spin direction
\footnote{For a muonic atom with a non-zero nuclear spin, it is known that
the residual muon spin polarisation at the 1$s$ state is
significantly reduced, but even in this case,
it could be recovered by using a spin-polarised
nuclear target \cite{Kadono:1986zz, Kuno:1987dp}.}.  For simplicity, we
will   in the following only discuss
the ten operators  that create
an $e_L$.

Notice that the conventions of eqn (\ref{CiriglianoN}) label
operator coefficients with the chirality of the muon,
which is opposite to the electron for dipole, scalar,
pseudoscalar and tensor operators.

\item  Finally, the operators
can also be divided into those that mediate
 SI or SD conversion.
 In the body of the
paper, we will discuss the SI rate, to which contribute
the dipole that we neglect, and  the vector and scalar on the
neutron and proton. These appear in the amplitude
weighted by  overlap integrals (see after eqn \ref{BRSIKKO}),
which are nucleus-dependent.  This suggests
that to constrain  the four operator
coefficients, one just needs to search for
$\mec$ on four sufficiently  different
targets.  (In order to measure the SI coefficients independently 
from SD ones, the targets could/should be chosen without 
SD contributions.)  

In the Appendix \ref{app:SD}, we
make some remarks on the SD rate, which
can be sensitive to six coefficients. However, 
quantitative calculations would require
nuclear matrix elements  that we did not find
in the literature.  

\een
 
\section{Targets as vectors, and the problem of  theoretical uncertainties}
\label{ssec:picture}

In a previous publication\cite{DKS},  a representation of targets
as vectors in coefficient space was introduced. The targets
are labelled by $Z$,  and  for SI transitions, 
the elements of the vector are the overlap integrals of KKO \cite{KKO}:
\beq
\vec{v}_Z = \left( \frac{D_Z}{4}, V^{(p)}_Z,S^{(p)}_Z,V^{(n)}_Z,S^{(n)}_Z \right) 
\label{v}
\eeq
The aim was to give a geometric, intuitive
measure of different targets ability to
distinguish coefficients.  
If the   operator coefficients are
  lined up in  a pair of vectors
labelled by the chirality of the outgoing
electron, such that for $e_L$:
\beq
\vec{C}_{L} = (\widetilde{C}_{D,R}, \widetilde{C}_{V,L}^{pp},\widetilde{C}_{S,R}^{pp},  \widetilde{C}_{V,L}^{nn},\widetilde{C}_{S,R}^{nn})
\label{vecC}
\eeq
(and similarly for $\vec{C}_{R}$), then 
the SI Branching Ratio  on target $Z$
(see eqn (\ref{BRSIKKO})
can be written  {\color{blue}
\beq
BR_{SI}  =B_Z{\Big [}  |\hat{v}_Z\cdot \vec{C}_{L} |^2 +  |\hat{v}_Z\cdot \vec{C}_{R} |^2 {\Big ]}   ~~~,
\label{ip2}
\eeq
where the  numerical value of the coefficient 
\begin{equation}
B_Z =  \frac{32G_F^2 m_\mu^5 |\vec{v}_Z|^2}{\Gamma_{cap}(Z)}
\label{BZ}
\end{equation}
is listed
in table \ref{tab:1} for some targets.}
If two target vectors are parallel, they probe
the same combination of couplings, and if they
are misaligned, they could allow to distinguish among
the coefficients.

To quantify  how ``misaligned''
targets   need to be, in order to
differentiate among coefficients,
we should take into account the  theoretical
uncertainties.
These are a significant complication,
because they make uncertain    which
combination of coefficients is constrained
by which target. To illustrate
the problem,
we suppose coefficient space is two-dimensional.
This allows to draw pictures.

\begin{figure}[ht]
\unitlength.5mm
\begin{center}
\epsfig{file=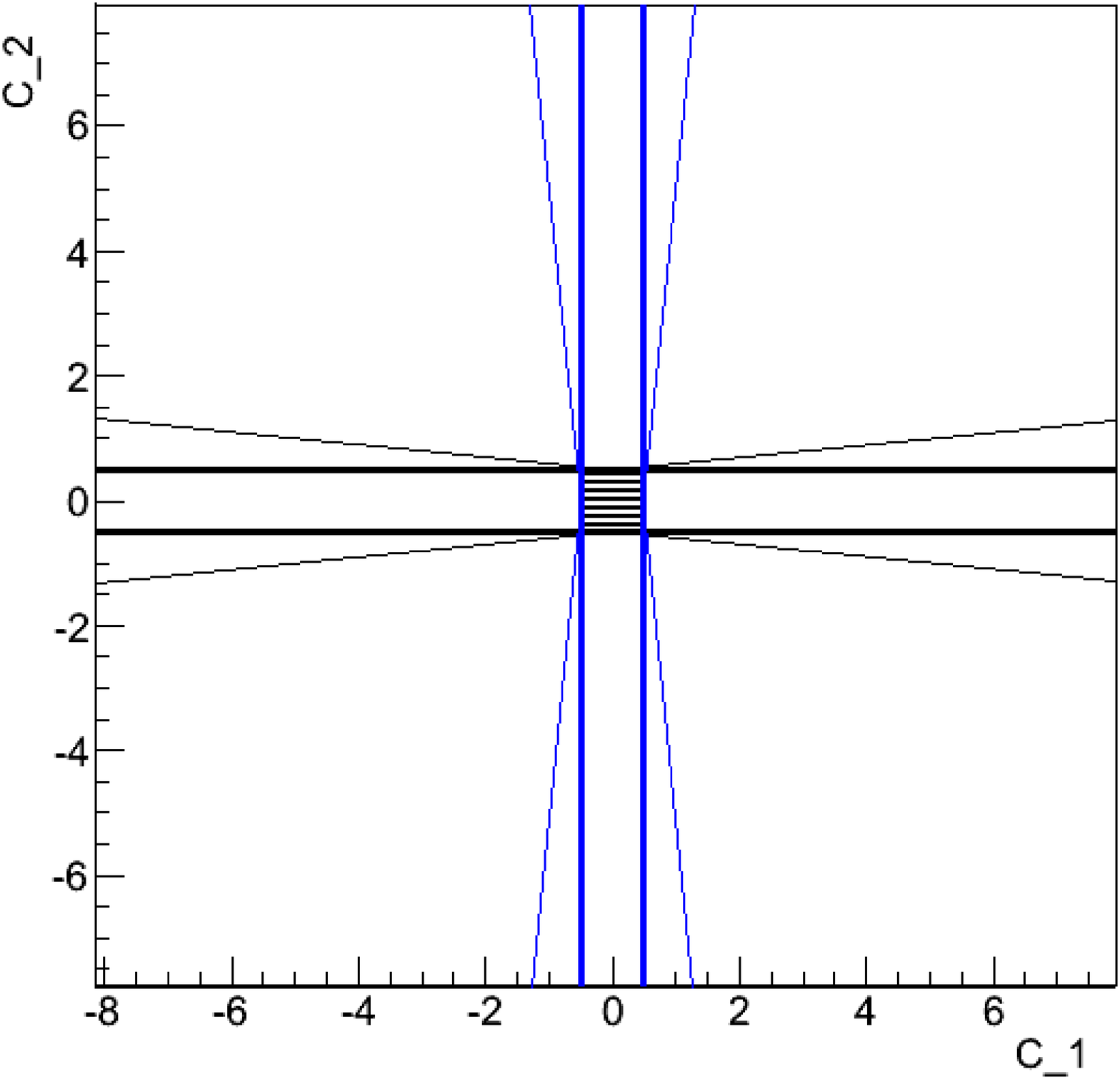,height=6cm,width=6cm}
\epsfig{file=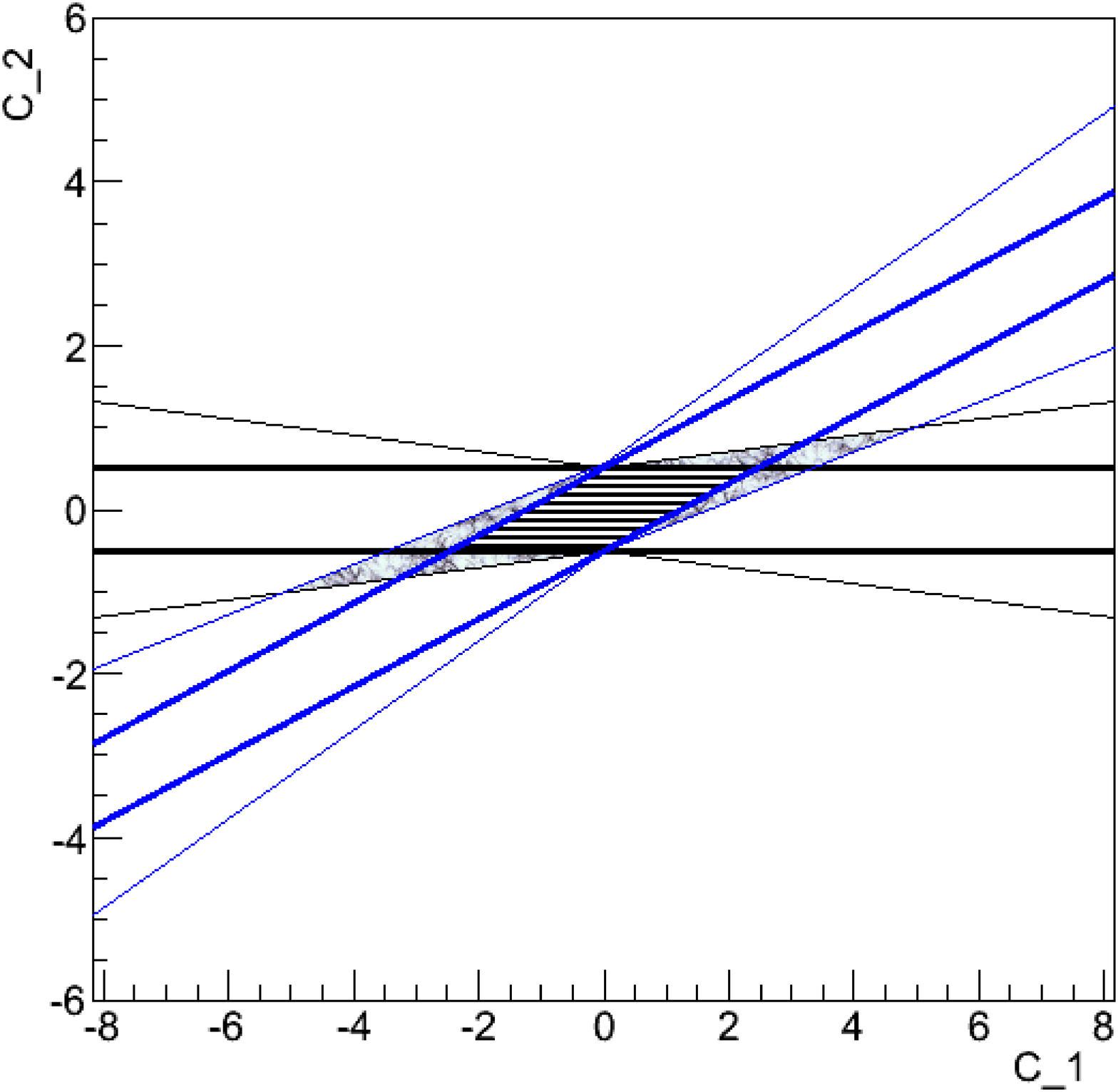,height=6cm,width=6cm}
\epsfig{file=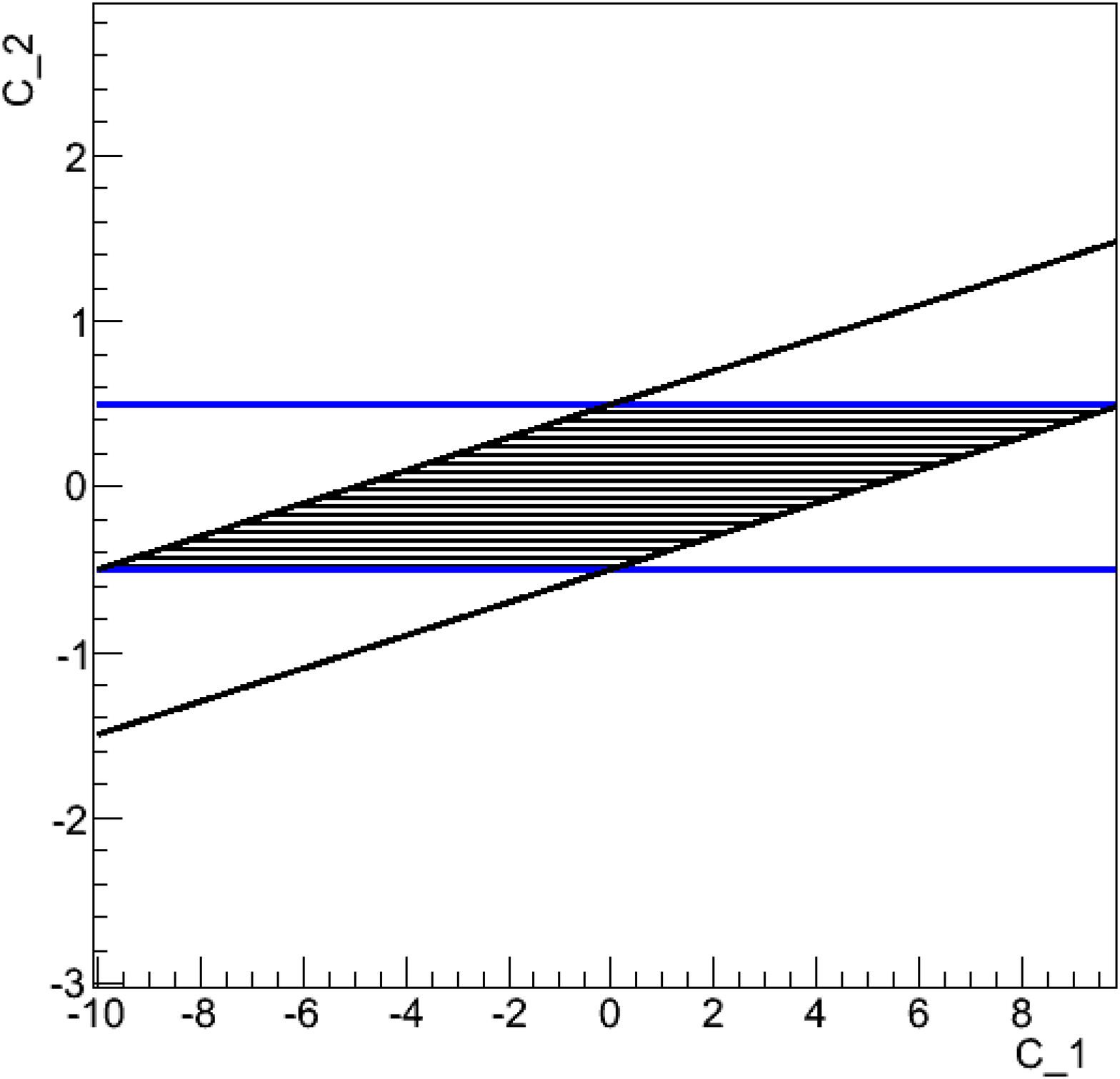,height=6cm,width=6cm}
\epsfig{file=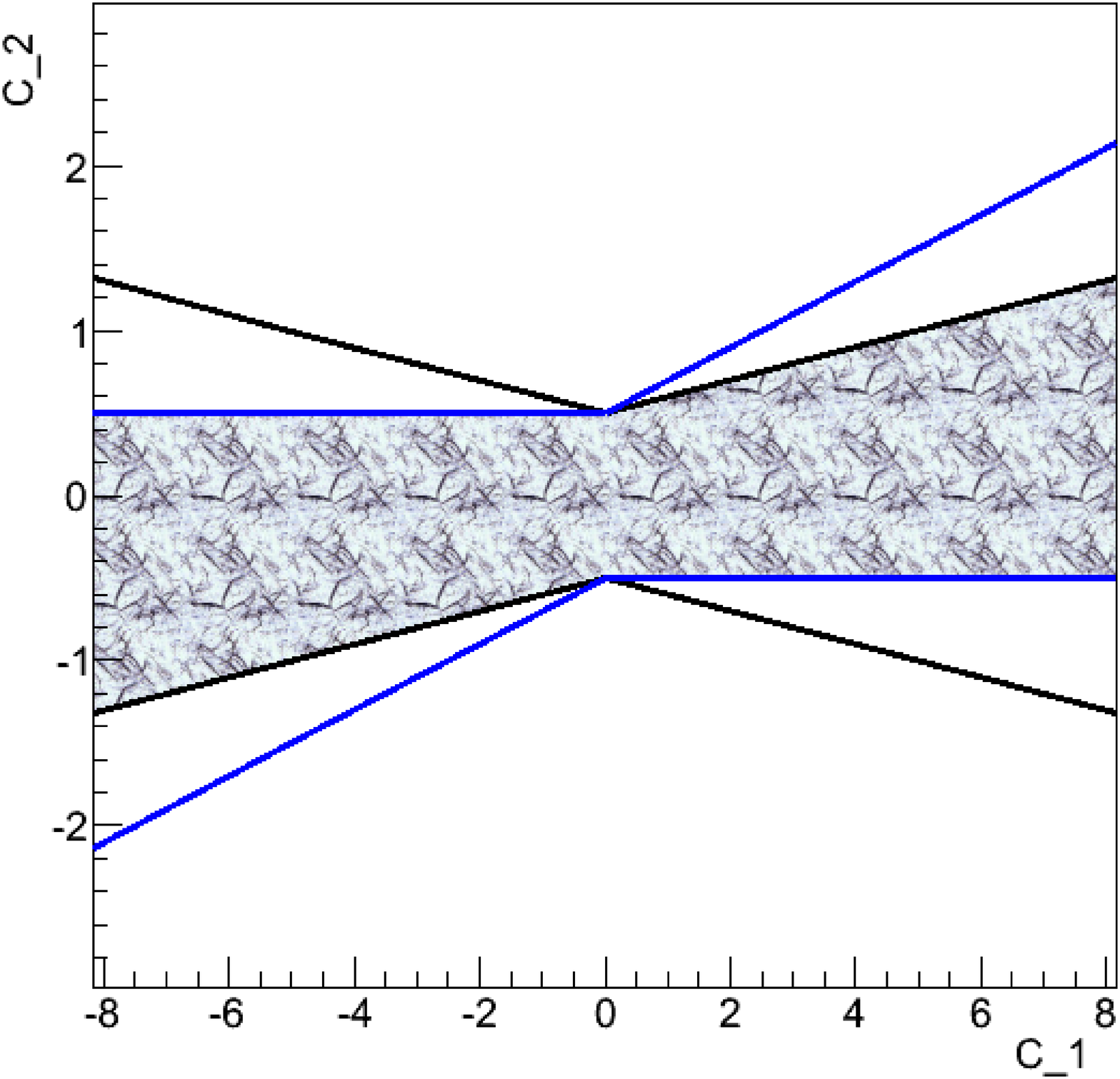,height=6cm,width=6cm}
\end{center}
\caption{ Illustration of the
impact of theoretical uncertainties  on
the determination of operator coefficients,
when combining results from two experiments. 
The allowed region neglecting
theory uncertainties is stripped; the larger
grey areas  are allowed when
 theoretical uncertainties are included.
The upper  left plots is for two experiments
that measure orthogonal parameters, the
upper right plot is for two experiments
who measure correlated parameters but
with manageable uncertainty, and
the lower two plots represent the case
where the two experiments do not give
independent information when the theory
uncertainty is included. 
\label{fig:draw} }
\end{figure}

If a first observable  $T_1$,
 can be computed with negligible theoretical uncertainty 
to   depend on $|C_1|^2$, and a second 
observable  $T_2$,
  similarly can  be computed to   depend on $|C_2|^2$,
then the values of  the coefficients
 respectively allowed by null results in the two experiments 
 are  inside the thick lines of   the top left plot in Figure \ref{fig:draw}.
 The  central stripped (dark)  region
 is allowed when the two experimental results are
 combined.  In reality, the allowed
 region should be more the shape of a circle,
since the experimental uncertainties are
(in part) statistical. However, we
neglect this detail because it is not
the subject of our discussion.

 Suppose now that there is some theoretical
 uncertainty $\epsilon$ in the calculations, such that
 $T_1$ depends on $|C_1 (1 \pm \epsilon)  \pm \epsilon C_2|^2$,
 and  $T_2$ depends on $|C_2 (1 \pm \epsilon) \pm \epsilon C_1|^2$.
Then provided $\epsilon \ll 1$ ($\epsilon \simeq \pi/32 \simeq .1$
in the Figure),
the regions respectively allowed by the two experiments
are  the bowties within the thin lines of the upper left
plot in Figure\ref{fig:draw}. The region allowed
 by the combined experiments is essentially unchanged
 (still the central square).

Consider next  a situation more relevant to $\mec$,
illustrated in the upper right plot of Figure \ref{fig:draw}.
The second observable $T_2$ again depends on
 $|C_2(1 \pm \epsilon) \pm \epsilon C_1|^2$,
 but  $T_1$  depends on
 $|\cos\theta C_2 - \sin \theta C_1 |$, where 
$\theta  \simeq  \pi/8 \pm \epsilon$.
Neglecting theoretical uncertainties,
the allowed regions for the two experiments
are respectively between the thick blue  lines,
and thick black lines. The stripped diamond is 
the parameter range  consistent with both 
experiments.  But if the theory uncertainty is
taken into account, the allowed regions of
the two experiments are  respectively
enclosed by the thin  blue and
black lines.  The region allowed by the
combined observations is the
grey diamond, which includes the stripped one. 
So we see that 
the  theoretical uncertainty   changes the allowed
region by factors of ${\mathcal O}(1)$.

Finally, in the lower two plots of
Figure  \ref{fig:draw},   $T_1$  depends on
 $|\cos\theta C_2 -\sin \theta C_1|$, where 
$\theta  \simeq  2\epsilon \pm \epsilon$.
If the theory uncertainty is neglected, as illustrated
in the lower left plot,  the region allowed by
the two experiments corresponds
to the stripped diamond.  
However, when
the angle uncertainty is taken into account,
both bars can be rotated towards each
other, such that they point in the same
direction, and  any value of $C_1$ is allowed.  This is
illustrated in the lower right plot,
where the allowed region is  grey, 
and gives no constraint on $C_1$.

The allowed range for $C_1$ would be
finite for 
\beq
\theta > 2 \epsilon
\eeq
which we take as the condition that
two observables  constrain independent
directions in coefficient space. (Recall that
$\theta$ is the angle between  the two
observables, represented as vectors in coefficient space,
and $\epsilon$ is the theoretical uncertainty
on the calculation of $\theta$).

For $\mec$, the theoretical  uncertainties in the
calculation of the  rate and  the overlap integrals
were discussed in  \cite{DKS}.
The current  uncertainties were
estimated as $\epsilon \lsim  10\, \%$.
This is based on KKO's estimate
of the uncertainties
in their overlap integrals, which is   $\lsim 5\,\%$ for light nuclei,
and  $\lsim 10\,\%$ for heavier nuclei, and on  NLO effects
in   $\chi$PT.
These are parametrically $10\,\%$,
 and could, for instance
 change the form of eqn (\ref{ip2}),
 as occurs in WIMP scattering \cite{Cirigliano:2012pq},
making it  impossible to parametrise
targets as vectors.
In the following section, we  take the current  
 uncertainties to be  $ \epsilon \sim 10\,\%$, or possibly
 $5\,\%$ for light targets,  implying that
 two targets can give  independent constraints 
 if they are misaligned by $\theta \gsim .2$
 (or possibly  $\theta \gsim .1$ for light targets).

\section{Comparing current bounds}
\label{sec:5}

In section \ref{sec:3}, it was suggested that the four
scalar and vector coefficients could be constrained   or
measured  by searching for $\mec$ in four ``sufficiently different''
targets. And we see from Table \ref{tab:1} that
there is data for Sulfur, Titanium, Copper, Gold and Lead.
However,  as estimated in the previous section,
``sufficiently different'' means misaligned by 10-20\%,
so in this section, we calculate the misalignment between  the
targets for which there is data.

Recall that  targets are described  by vectors (see eqn \ref{v}),
that live in the same space as the operator coefficients.
However, the components of the target vectors   are all  positive,
meaning the misalignment angle between  any two target
vectors is  $<\pi/2$, or equivalently, that the  target
vectors point all in the first quadrant.

\begin{figure}[ht]
\unitlength.5mm
\begin{center}
\epsfig{file=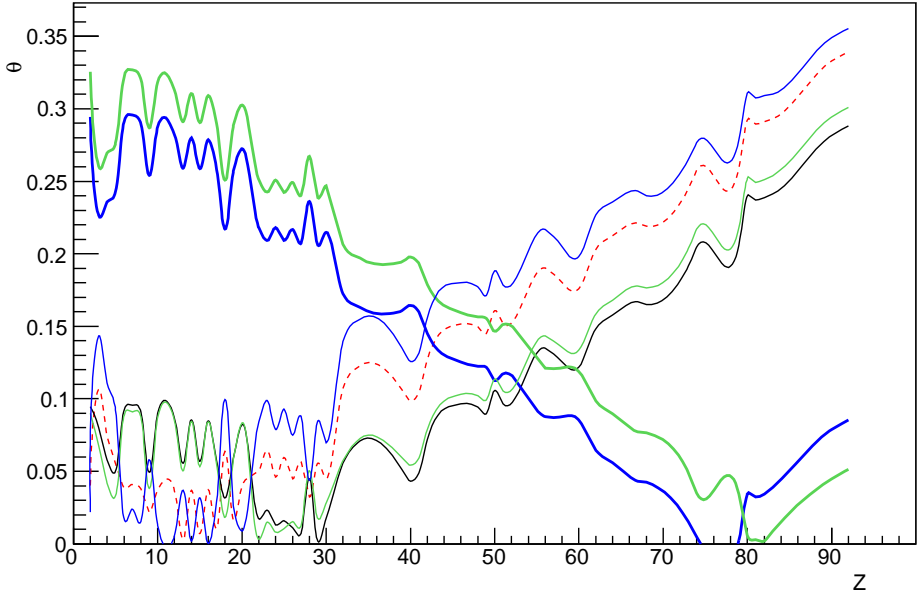,height=9cm,width=14cm}
\end{center}
\caption{
Angle $\theta$ between a target vector (eg dashed red = Aluminium) and
other targets labelled by $Z$. The angle is obtained
as in eqn (\ref{thetadefn}), with
all the dipole coefficients  set to zero.
The solid lines represent the targets for
which there is currently data (see table \ref{tab:1}).
From smallest to largest value of $\theta$ at large $Z$,
they are: thick green =  Lead, thick blue = Gold,
black = Copper, thin green = Titanium,
 dashed red = Aluminium,
and thin blue is  Sulfur.  We assume that two
targets can probe different coefficients
if their misalignment angle is
$\theta \gsim  0.2 $ radians (or 0.1). 
\label{fig:AlD} }
\end{figure}

The  angle between target $T$ and target $Z$
can be estimated from the
normalised inner product
\beq
\frac{\vec{v}_Z\cdot \vec{v}_{T}}{|\vec{v}_Z|| \vec{v}_{T}|} \simeq \cos \theta
\simeq 1 -\frac{\theta^2}{2}
\label{thetadefn}
\eeq
In Figure \ref{fig:AlD} are plotted the
misalignment angles\footnote{Since
the  current MEG bound on the dipole coefficients
constrains them to be  below the sensitivity
of the current $\mec$ bounds, the
dipole overlap integral was set to
zero in  obtaining this Figure.} between the targets
of table \ref{tab:1}, and the other possibilities
given by KKO, labelled by $Z$. 
The thin blue line  in 
Figure \ref{fig:AlD}
(the line with largest $\theta$ at high $Z$)
is the misalignment angle with respect
to Sulfur, and
the thin green line
(the  solid line with  the second largest $\theta$ at high $Z$)
is the misalignment angle with respect
to Titanium. 
So the blue
line at Z=22 (Titanium) is  equal
to the green line at 16 (Sulfur),
and both give  $\theta \sim 0.08$
between Sulfur and Titanium,
suggesting that these  constrain
the same  combination of coefficients. 
On the other hand,   Gold
probes different coefficients from
the light targets (as anticipated by KKO \cite{KKO}),
but  Gold and Lead cannot distinguish
coefficients.  Also  Copper and Titanium
 do not give independent constraints.
So the current experimental bounds on
$\mec$   constrain two  combinations
of the four coefficients    present in
$\vec{C}_{L}$ (similarly,  two combinations
in $\vec{C}_{R}$). Thus, the current experimental
bounds can be taken as the SINDRUM-II constraints 
from Titanium and Gold.

{\color{blue}
These two experimental bounds constrain coefficients
in the two-dimensional space spanned by
 $\hat{v}_{Ti}$ and  $\hat{v}_{Au}$. 
The   bounds can be taken
to apply to $\vec{C} \cdot \hat{v}_{Ti}$
and to  $\vec{C} \cdot \hat{v}_{\perp}$, where
$\vec{v}_\perp$ is  component of the Gold target vector orthogonal
to  $\hat{v}_{Ti}$:
\beq
\hat{v}_\perp \equiv  \frac{\hat{v}_{Au} - \cos \phi \hat{v}_{Ti}}{\sin \phi}
\eeq
and  $\phi$ is the angle between Gold and Titanium,
so $\cos \phi = \hat{v}_{Ti} \cdot \hat{v}_{Au}$,
and $\sin \phi = 0.218$.
The allowed  values of the coefficients
satisfy 
\bea
BR_{Ti}\equiv BR(\mu Ti  \to e Ti ) &=& 234|\vec{C}\cdot  \hat{v}_{Ti}|^2 <
BR^{exp}_{Ti}\equiv 4.2\times 10^{-12}
\nonumber\\
BR_{Au} \equiv BR(\mu Au  \to eAu ) &=& 285 | \cos\phi (\vec{C}\cdot  \hat{v}_{Ti})
+ \sin \phi (\vec{C}\cdot  \hat{v}_\perp) |^2 <
BR^{exp}_{Au }\equiv 7.0\times 10^{-13}~~~.
\eea
We can construct a covariance matrix  $V$, whose diagonal
elements will be   the constraints on $|\vec{C}\cdot  \hat{v}_{Ti}|^2$
and $|\vec{C}\cdot  \hat{v}_\perp|^2$, as
\beq
\vec{C} \cdot V^{-1} \cdot \vec{C}^T =
\frac{BR_{Ti}}{BR_{Ti}^{exp}}
+ \frac{BR_{Au}}{BR_{Au}^{exp}}
\label{covar}
\eeq
which gives
\bea
|\vec{C}\cdot  \hat{v}_{Ti}|^2 &\leq & \frac{BR^{exp}_{Ti}}{B_{Ti}} = 1.8\times 10^{-14}\\
|\vec{C}\cdot  \hat{v}_\perp|^2 &\leq &
\frac{BR^{exp}_{Au}}{B_{Au}}\frac{1}{\sin^2\phi}
+ \frac{BR^{exp}_{Ti}}{B_{Ti}}\frac{\cos^2\phi}{\sin^2\phi}
= 0.44 \times 10^{-12} 
\label{boundsnow}
\eea
These bounds can be expressed  in terms 
of quark operator coefficients at a higher scale
by matching the nucleon operators
onto quark operators, and running the coefficients up with Renormalisation
Group Equations(RGEs).
This matching and mixing process ensures that
almost all $\mu \to e$ flavour-changing
operators   at  the scale   $m_W$  will  contribute to
$\mec$ at tree or one-loop  order. We give an
example in eqn (\ref{RGEs}).

}

\section{Selecting future targets}
\label{sec:target}

The upcoming
COMET and Mu2e experiments plan to use an Aluminium target,
illustrated as a red dashed line in Figure \ref{fig:AlD}.
Unfortunately, it is  only  misaligned with respect to Titanium
and Sulfur  by  a few percent,
so with current theoretical uncertainties,
Aluminium probes  the same
combination of  SI coefficients as  
Titanium (and Sulfur).

\begin{figure}[ht]
\unitlength.5mm
\begin{center}
\epsfig{file=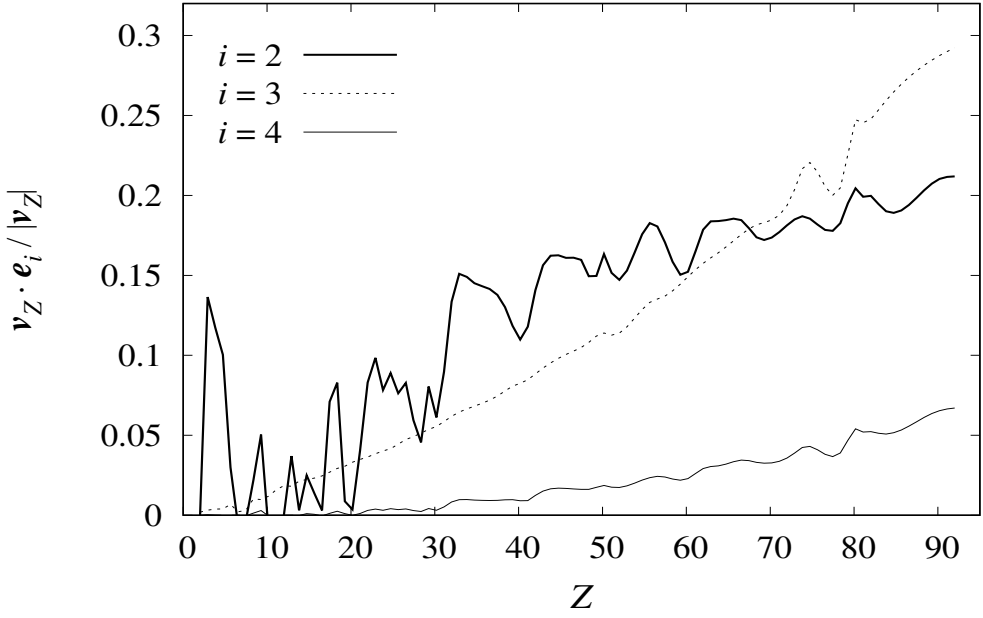,height=9cm,width=14cm}
\end{center}
\caption{Projections of normalised  target vectors
$\hat{v}_Z = \vec{v}_Z/| \vec{v}_Z|$  onto
the unit vectors of eqn \ref{ei}.  The jagged  thick 
line is the projection onto  $\hat{e}_2$ and parametrises 
  the targets sensitivity to the $n$ to $p$ difference,
the dotted black line is the projection onto $\hat{e}_3$ which 
parametrises the sensitivity  to  the scalar-vector difference,
and  the   thin 
line is the projection onto
the remaining  direction.
\label{fig:3} }
\end{figure}

It is therefore  interesting to explore which  targets
could   measure a different  combinations of
coefficients from Aluminium. As noted by KKO,
the scalar and vector overlap integrals grow
differently with $Z$, and  using
targets  with different neutron to proton ratios  could
allow to differentiate   coefficients on protons
from those on neutrons.  To quantify these
differences, we introduce  four orthonormal vectors
in  the space of nucleon overlap integrals:
\bea
\hat{e}_1 &=&\frac{1}{2} (1, 1, 1, 1)\nonumber\\
\hat{e}_2 &=&\frac{1}{2} (-1,-1,1,1)\nonumber\\
\hat{e}_3 &=&\frac{1}{2} (1,-1,1,-1)\nonumber\\
\hat{e}_4 &=&\frac{1}{2} (-1,1,1,-1)\label{ei}
\eea
 Dotted into the coefficients, $\hat{e_1}$ measures the
 sum of coefficients,  $\hat{e_2}$  is the difference
 between  coefficients on protons and neutrons,
 $\hat{e_3}$ is  the vector - scalar difference,
 and $\hat{e_4}$ is the remaining direction.
All the targets are mostly aligned on $\hat{e}_1$;
this is expected as the overlap integrals are
 of comparable size, and all positive
\footnote{One way to see this, is to project
the target vectors onto the basis of eqn \ref{ei}.
We find that 
$\vec{v}_Z \cdot \hat{e_1} \geq 0.93 |\vec{v}_Z|$
 for all $Z$, so we do not plot the projection
 onto $\hat{e}_1$. It decreases  with Z.}.
Indeed, for Aluminium,  the target vector
$\vec{v}_{Al}$  and  $\hat{e}_1$ are almost parallel: 
 $\vec{v}_{Al} \cdot \hat{e_1} \geq 0.997 |\vec{v}_{Al}|$.
So we suppose that this  sum
 of coefficients   is measured on Aluminium, 
  and  plot in Figure \ref{fig:3} the  projection
of the target vectors 
onto $\hat{e}_2$(thick, continuous line), $\hat{e}_3$ (dashed) and $\hat{e}_4 $
(thin).

Figure \ref{fig:3} shows  that comparing heavy to light targets
can distinguish scalar vs vector coefficients
(or constrain both in the absence of a signal).
The neutron to proton ratio also increases
with atomic number, but perhaps a more promising target for
 making this difference would be  Lithium,
 with four neutrons and three protons:  the
 theoretical uncertainties could be more manageable,
 and  the scalar-vector difference is suppressed.
 In addition, being light, it has a long lifetime,
 making it appropriate for  the COMET and Mu2e experiments.

Unfortunately, it seems that $\mec$ targets
have little sensitivity to $\hat{e}_4$,  which
{\color{red} measures some isospin-violating
difference between scalars and vectors.}

{\color{blue}

To illustrate the bounds that could be obtained in the
future with COMET or Mu2e, we suppose that $\mec$ is not observed
on Lithium($Z=3$) or Aluminium($Z=13$) at
Branching Ratios $ BR^{exp}_{Li}, BR^{exp}_{Al} \ll BR^{exp}_{Au}$.
We  write  the target vectors as
\bea
\hat{v}_{Al}&\approx & \hat{e}_1 \nonumber\\
\hat{v}_{Li}& = & (\hat{v}_{Li}\cdot \hat{e}_1)  \hat{e}_1 +
(\hat{v}_{Li}\cdot \hat{e}_2) \hat{e}_2\nonumber\\
\hat{v}_{Au}& =  & (\hat{v}_{Au}\cdot \hat{e}_1) \hat{e}_1 +
(\hat{v}_{Au}\cdot \hat{e}_2) \hat{e}_2
+ (\hat{v}_{Au}\cdot \hat{e}_3)  \hat{e}_3 ~~~.
\nonumber
\eea
 Then a  $3\times 3$ covariance matrix
can be
 obtained by combining the experimental upper bounds  as in equation (\ref{covar}), which gives the constraints
 \bea
|\vec{C}\cdot  \hat{e}_1|^2 &\leq & \frac{BR^{exp}_{Al}}{B_{Al}}
\label{boundstobe1}\\
|\vec{C}\cdot  \hat{e}_{2}|^2 &\leq &
\frac{BR^{exp}_{Li}}{B_{Li}}\frac{1}{|\hat{v}_{Li}\cdot \hat{e}_2|^2}
+ \frac{BR^{exp}_{Al}}{B_{Al}} \frac{|\hat{v}_{Li}\cdot \hat{e}_1|^2}{|\hat{v}_{Li}\cdot \hat{e}_2|^2}
\nonumber\\
|\vec{C}\cdot  \hat{e}_3|^2 &\lsim & \frac{BR^{exp}_{Au}}{B_{Au} }
\frac{1}{|\hat{v}_{Au}\cdot \hat{e}_3|^2}
\label{boundstobe}
\eea
where $|\hat{v}_{Li}\cdot \hat{e}_2| = 0.142$, 
$|\hat{v}_{Au}\cdot \hat{e}_3| =0.217$,
and  terms were neglected in the
bound on $|\vec{C}\cdot  \hat{e}_3|$, assuming that
 $ BR^{exp}_{Li}, BR^{exp}_{Al} \ll BR^{exp}_{Au}$.
 
As mentioned at the  end of the previous  section,
these bounds can be expressed in terms of coefficients
of quark operators at some higher scale (for instance $m_W$
or the New Physics scale) by matching the nucleon coefficients
to the quark coefficients at 2 GeV  as $\widetilde{C}^{NN}_{O,X} =
\sum_q G_O^{Nq} C_{O,X}^{qq}$ (where the $\{G_O^{Nq}\}$ are tabulated
for instance in \cite{DKS}), then expressing
the  coefficients at 2 GeV in terms of
the high scale coefficients using the RGEs  (see,{\it eg} \cite{CDK,DKS,PSI}).
The  constraint of eqn (\ref{boundstobe1}) can be
approximated as
\bea
\sqrt{\frac{BR^{exp}_{Al}}{33}}&\gsim& {\Big|}
3C^{uu}_{V,L} + 3C^{dd}_{V,L} + 11C^{uu}_{S,R} +
 11C^{dd}_{S,R} +  0.84C^{ss}_{S,R}  
+ \frac{4m_N}{27m_c}C^{cc}_{S,R} + \frac{4m_N}{27m_b}C^{bb}_{S,R}
{\Big|} \nonumber \\
&\gsim &  {\Big|}
3C^{uu}_{V,L} + 3C^{dd}_{V,L} 
+\frac{\alpha}{\pi}
{\Big [}3C^{dd}_{A,L}  -6C^{uu}_{A,L} {\Big]}\log
+ \frac{\alpha}{3\pi}
[ C^{ee}_{V,L} + C^{\mu \mu}_{V,L}] \log
- \frac{\alpha}{3\pi}
[ C^{ee}_{A,L} + C^{\mu \mu}_{A,L}] \log
\nonumber\\ 
&&  - \frac{2\alpha}{3\pi}
{\Big [}2( C^{uu}_{V,L} + C^{cc}_{V,L})
-( C^{dd}_{V,L} + C^{ss}_{V,L}  + C^{bb}_{V,L})
-( C^{ee}_{V,L} + C^{\mu \mu}_{V,L}  + C^{\tau \tau}_{V,L}) {\Big ]}\log
\nonumber\\
&&
+  \lambda^{a_S} {\Big (}11 C^{uu}_{S,R} +
 11C^{dd}_{S,R} +  0.84C^{ss}_{S,R}
 + \frac{4m_N}{27m_c}C^{cc}_{S,R} + \frac{4m_N}{27m_b}C^{bb}_{S,R} {\Big )} 
\nonumber\\ 
&&
+ \lambda^{a_S} \frac{ \alpha}{ \pi} 
{\Big [} \frac{13}{6 }  (11 C^{uu}_{S,R}   + \frac{4m_N}{27m_c} C^{cc}_{S,R} )
+ \frac{5 }{3}( 11 C^{dd}_{S,R} + 0.84C^{ss}_{S,R} + \frac{4m_N}{27m_b}C^{bb}_{S,R}){\Big ]} \log
\nonumber\\ 
&&
 -\lambda^{a_T}f_{TS}  \frac{4\alpha}{ \pi}  
{\Big[} 22 C^{uu}_{T,RR}   + \frac{8m_N}{27m_c} C^{cc}_{T,RR} 
- 11 C^{dd}_{T,RR} - 0.84C^{ss}_{T,RR}  -\frac{4m_N}{27m_b}C^{bb}_{T,RR}) {\Big ]} \log
{\Big|} \label{RGEs}
\eea
where in the first  inequality, the quark coefficients are
at the scale of 2 GeV, and we used the $\{G_S^{Nq}\}$ from
the lattice \cite{lattice}. In the second inequality,
the coefficients are at $m_W$ (we suppress the dependence on scale to
avoid cluttering the equations),
$\log \equiv \log ({m_W}/2{\rm GeV}) \simeq 3.7$,
$\lambda = \alpha_s(m_W)/\alpha_s(2{\rm GeV}) \simeq 0.44$,
$f_{TS} = 23(\lambda^{-16/23} - \lambda)/39/(1-\lambda)\simeq 1.45$, 
and $a_S = -12/23, a_T = 4/23 $. In both  the first and second
expressions, the  tree-level tensor contribution to
SI $\mec$ is neglected because it is smaller than
the loop mixing into the scalar, and the scalar top
 coefficient $C_{S,R}^{tt}$  is also neglected.}

{\color{blue}
\section{Flat Directions and Tuning in EFTs}
\label{sec:FlatDir}

We now want to  compare
 the  number of  constraints on $\mu \to e$
 flavour-changing coefficients, to 
the number of operators
in a  ``complete'' basis
--- the difference  will be
the number of ``flat'' or unconstrained directions in
parameter space.

In this counting, it is important to distinguish
 constraints from sensitivities. The bounds given
 in eqn (\ref{boundsnow}) are constraints, meaning  that
 the sum on the left cannot exceed the number on the right.
 This is different from the commonly-quoted sensitivities
 (or one-operator-at-a-time bounds), which give the value of a
 coefficient below which it is unobservable, and  which
 do not allow for the possibility of cancellations
 among coefficients. 

In the Effective Theory below $m_W$,
a useful basis is the set of operators
that  are  QED and QCD invariant, 
and     that  describe  all  the
 three or four-point functions that  change
 lepton flavour from $\mu \to e$.
 These operators are of dimensions
 five, six and seven, and are 
 listed, for instance, in  \cite{megmW}. 
We restrict to  four-fermion operators
 whose second fermion bilinear is quark
 or lepton flavour-conserving
(only these can contribute to $\mec$), 
 in which case the basis contains 82 operators.

Eqn (\ref{boundsnow}) gives the  current  $\mec$ bounds
 on four combinations of SI coefficients
(two bounds for each electron chirality).
There  also should be two constraints
on  proton Spin-Dependent  coefficients  from  Gold
(since it has 79 protons), however
the rate  has not been calculated and
could be quite small. In
addition, there could be
 two  constraints
 on  neutron Spin-Dependent coefficients
 from Titanium, if the
 experimental target  contained
isotopes with  an odd number of neutrons. 
So current data gives six or eight bounds. 

 With a wider variety of targets, and
 improved theoretical calculations,
 we  showed in eqn (\ref{boundstobe})
that it could be possible to constrain
 6 of the 8 SI coefficients, and
 we argued that eight
 of the 12 SD coefficients could
 be constrained. 

There are also stringent constraints on 
$\mu \to e$ flavour-changing operator coefficients
from $\meg$ and $\meee$.  Constructing a
covariance matrix for these two processes
using the theoretical Branching Ratio
formulae from \cite{KO} gives the bounds:
 \bea
|C_{D,R}|^2 ,|C_{D,L}|^2  &\leq&
\frac{BR^{exp}_{\meg} BR^{exp}_{\meee}}{ 205e^2 BR^{exp}_{\meg} + 384\pi^2 BR^{exp}_{\meee}} \nonumber \\
|C_{S,RR}|^2,|C_{S,LL}|^2&<& 8  BR^{exp}_{\meee} \nonumber\\
|C_{V,RR}|^2 ,|C_{V,LL}|^2  &\leq& \frac{ BR^{exp}_{\meee}}{ 2}\left(
1 +  \frac{32e^2BR^{exp}_{\meg}}{ 205e^2BR^{exp}_{\meg} +384\pi^2 BR^{exp}_{\meee}}\right)\nonumber \\
|C_{V,RL}|^2 ,|C_{V,LR}|^2 & \leq & BR^{exp}_{\meee}
\left( 1 +  \frac{16e^2BR^{exp}_{\meg}}{ 205e^2BR^{exp}_{\meg} +384\pi^2 BR^{exp}_{\meee}}\right)
\label{bdmegmeee}
\eea
where the four-lepton operators are $
{\cal O}_{V,XY} = (\overline{e}\gamma^{\a} P_X \mu ) (\overline{e} \g_\a P_Y e)$, ${\cal O}_{S,XY} = (\overline{e} P_X \mu ) (\overline{e}  P_Y e)$.

Combining the constraints from $\meg$, $\meee$
and  the current $\mec$ data,  gives  14 to 16
constraints. In the  future, $\mec$ could give
two  more SI constraints and four more SD bounds,
for a total of 22. 
Each constraint applies to an
lengthly linear combination of
 coefficients at $m_W$; nonetheless,
there are therefore 60  to 68 
combinations of coefficients (in the
basis of QED$\times$QCD-invariant operators
discussed above) which are
unconstrained by $\mec$, $\meg$ and $\meee$.

In order to constrain the multitude 
of flat directions, other processes can be
considered, such as  contact interaction searches
at the LHC \cite{CS} or vector meson
decays \cite{Upsilon}. However, it might be
difficult to   find a sufficient number of
restrictive constraints. Let us here speculate about
how credible it is for  a model to sit out along
such a ''flat'' direction,  where the operator coefficients
must be tuned against each other. We suppose
that the coefficients
parametrise the low energy behaviour
of a renormalisable and natural high-scale model. 
However, since the coefficients are unknown functions
of the model parameters, cancellations that
reflect  symmetries of the model  
could appear fortuitous in the  EFT.  We
therefore allow arbitrary cancellations among
coefficients  at the high scale. 
Then   we assume that the  model
 cannot know the scale at
which we do experiments (despite that
this is determined by mass ratios which
it does know), so we do not allow
coefficients to cancel the logarithms
from Renormalisation Group running. 

We see from eqn  (\ref{RGEs}) that the scalar and tensor
operators run significantly with QCD, which suggests
that they cannot
cancel to more than one significant
figure against  each other or vector/axial  coefficients.
So a single constraint such as eqn  (\ref{RGEs}), naturally
implies three  independent constraints on
the vectors, scalars and tensors.
Then within each subset of operators, the
QED anomalous dimension matrix could be
diagonalised, with,  in general,
non-degenerate anomalous dimensions. so
 coefficients that cancel  at the
high scale, may  differ by   ${\cal O} (\frac{\alpha}{4\pi} \log)$
 at low energy. We therefore conclude that
 cancellations  are only ``natural''
 to  3  significant figures, among operator
 coefficients with unequal anomalous dimensions: if the constraint
 applies to a sum of operator coefficients $C_j$  weighted by numbers $n_j$
\beq
|\Sigma_j n_j C_j| < \epsilon
\eeq
then each $C_j$ in the sum satisfies a ``naturalness bound'' of order
\beq
C_j \lsim \frac{4 \pi \epsilon}{\alpha n_j} ~~~.
\label{EFTnatural}
\eeq

}

\section{Summary}
\label{sum}

This letter studies the  selection
of targets for $\mec$,  with the aim  that they probe
independent combinations of  $\mu \to e$ flavour-changing parameters,
while including  the  theoretical uncertainties
 of the calculation. The rate is parametrised
via the operators given in eqn (\ref{CiriglianoN}),
and the theoretical uncertainties are reviewed in section
\ref{ssec:picture}. We take the current uncertainties
to be $\sim 10\%$,  and anticipate that this could be
reduced to 5\% in the future.

Using a parametrisation of targets as vectors
 in the space of operator coefficients,  we reproduce
 the observation of Kitano, Koike and Okada (KKO) \cite{KKO},
 that comparing light to heavy targets allows to distinguish 
 scalar from vector operator coefficients. We
 also observe that  comparing light targets with
 very different neutron-to-proton ratios could allow
 to distinguish operators involving neutrons from those
 involving protons. A reduction in the
 theory uncertainty would help to  make this distinction.
 Lithium is the most promising target
  in  the list for which KKO computed
 overlap integrals,  however other light isotopes
 with higher $n/p$ ratios, such as
 Beryllium10, could be interesting to consider.

The Spin-Dependent (SD) conversion rate is mentioned
in the Appendix. We reiterate that the  neutron  vs proton
operators can be  distinguished by
searching for  SD conversion on nuclei
with an odd  neutron  and   with an
odd proton.  Comparing the  SD rate on light
vs heavy nuclei could allow to distinguish
axial from tensor coefficients, but dedicated nuclear calculations
would be required to confirm this. 

{\color{blue}
We conclude that $\mec$  currently can constrain
six  to eight independent combinations of operator
coefficients, and in the future could constrain
fourteen coefficients. 
Combined with the eight bounds that
can be obtained from $\meg$ and $\meee$,
this gives 14 (now) to 22 (in the future)
constraints on the 82 operators in a QED$\times$QCD
invariant operator basis below $m_W$,  
so there  remain  60-68 ``flat directions'',
or combinations of coefficients that are unconstrained
by the data.

For a model to be situated  along one of the many flat directions,
requires cancellations among various operator coefficients.
We argued that it would be ``unnatural'' to have cancellations
between terms at different order in the $\alpha\log$ expansion
of EFT, so coefficients can only cancel against each other up to
${\cal O }(\frac{\alpha}{4\pi} \log)$,   and  coefficients whose sum
is constrained to be $\lsim \epsilon$,   should individually 
satisfy the ``natural'' bound $C\lsim  4 \pi \epsilon/\alpha$.


}

\section*{Acknowledgments}

S.D. thanks  the Galileo Galilei Institute for Theoretical Physics
for hospitality, and the INFN for partial support,
during the completion of this work.
This work is supported in part by the Japan Society for the Promotion of Science (JSPS) KAKENHI Grant Nos.   25000004 and 18H05231 (Y.K.),
and  Grant-in-Aid for
Scientific Research Numbers 16K05325 and 16K17693 (M.Y.).

\appendix

\section{Appendix: the SD contribution}
\label{app:SD}

In this Appendix, we discuss how different targets
could distinguish among the many operators that
contribute to  SD conversion.

We first recall the operators that
contribute to the SD rate. 
For a fixed electron  chirality,
the pseudoscalar, axial vector and tensor nucleon currents
become, in the non-relativistic limit~\cite{CdNP}
\bea
\overline{u}_N(p_f) \g_5 u_N(p_i)& \to&  \vec{q}\cdot \vec{S}_N/m_N
\nonumber\\
\overline{u}_N(p_f) \g^\a \g_5 u_N(p_i)&\to & (\vec{P}\cdot \vec{S}_N,
2 \vec{S}_N)/m_N
\nonumber\\
\overline{u}_N(p_f) \sigma^{jk}  u_N(p_i)&\to & 2\epsilon^{jkl} S_l
\nonumber\\
\overline{u}_N(p_f) \sigma^{0l}  u_N(p_i)&\to & (iq^l -2 P_j S_k \epsilon^{jkl})/2m_N 
\eea
where $q = p_i - p_f$, and $P = p_f + p_i$.
So they all connect the lepton current to the
spin of the nucleon, and 
at zero-momentum transfer, the pseudoscalar nucleon current vanishes  and
the  tensor  current is twice the axial current.
{\color{blue}
At finite momentum transfer ($q^2 \neq 0$),  the $P,A$ and $T$ operators
have different behaviours. Only the axial operator has been
studied at  $q^2 \neq 0$,  in the case
of  (spin-dependent) WIMP scattering.  Reference \cite{DKS}
made the curious observation for light targets, that the $q^2 \neq 0$
suppression of the vector and axial currents was very similar.
We use this  numerology to estimate the $S_I(m_\mu)/S_I(0)$
correction for Titanium and Gold in Table \ref{tab:1}.
As discussed in \cite{DKS}, this approximation may be
reasonable for light nuclei, but is incredible for heavy nuclei such
as Gold, where the muon wavefunction  could give additional suppression.

}

The SD  coefficients on neutrons, can be distinguished 
from those on protons,  by comparing targets with an
odd number of protons  or neutrons \cite{CDK}. This can be seen
from Table \ref{tab:1}, where  the spin
of a nucleus is largely due to the spin of the
one unpaired nucleon. For
instance, searching for $\mec$ on Aluminium,
and on  a Titanium target containing  a sufficient abundance of spin-carrying
isotope, would give independent constraints on
SD coefficients on the proton and neutron.

It is possible that comparing heavy and light
targets could  distinguish axial from tensor operators.  
The estimate for the  SD Branching ratio  
given in  eqn (\ref{BRCDK})  assumes light nuclear targets
(where the muon wavefunction is broader than the atom),
and exhibits a  degeneracy between the tensor and axial coefficients. 
If the same light-nucleus  approximation  is used
to compute the SI rate, 
then the  scalar  and vector overlap integrals would be the same.
Indeed, as pointed out by KKO, the scalar  overlap integral
becomes different from the vector in  heavy nuclei,
because   the negative energy component of the electron wavefunction 
becomes relevant, and has opposite sign for vector
and scalar  (see KKO, eqns 20-23). There is a similar sign difference
between tensor and axial operators, so one could
hope to distinguish tensor from axial operators
by comparing the SD rate in light and heavy nuclei.
For instance, table \ref{tab:1} suggests  that $\mec$ on gold
and lead could  distinguish the axial  from tensor operators
respectively for protons and neutrons.
However, one difficulty is that the SD rate is relatively
suppressed with respect to the SI rate by a factor $\sim 1/A^2$,
which becomes more significant  for heavier nuclei. The
second difficulty is that dedicated nuclear calculations
of the  expectation value  in the nucleus of the
various SD operators, weighted by  the lepton
wavefunctions, would be required. These calculations
currently do not exist.

Finally, in order to  be sensitive to the Pseudoscalar
operator,  and to obtain reliable predictions for
the SD rates, the finite momentum transfer should be taken into
account. However, then in  squaring the matrix element,
the spin sums do not factorise from the  sum of operator coefficients
(as occurs for the SI rate, see eqn \ref{BRSIKKO}).
This suggests that  the nuclear calculation would need
to be performed in the presence of the A,P and T operators
in order to  explore the prospects of distinguishing the pseudoscalar.

\end{document}